\documentclass[12pt]{article}
\usepackage{multicol}
\usepackage{subfigure}
\usepackage{amsmath}
\usepackage{amssymb}
\usepackage{amsfonts}
\usepackage{graphicx}
\usepackage{url}
\usepackage{ccaption}
\usepackage{booktabs} %
\usepackage{color}

\def\lv{LabVIEW}

\begin{document}

\title{A \lv{} Code for PolSK encoding}




\author{
Ram Soorat,  K. Madhuri\thanks{Present address, Department of Electrical and Computer Engineering, Southern Polytechnic State University,   United States of America}  and  Ashok Vudayagiri \\School of Physics, University of Hyderabad, \\ Hyderabad, 500046 {\bf India}}


\maketitle

\abstract{ We have developed an integrated software module for use in free space Optical communication using Polarization Shift Keying. The module provides options to  read the data to be transmitted from a file, convert this data to on/off code for laser diodes as well as measure the state of polarization of the received optical pulses. The Software bundle consists of separate transmitter and receiver components. The entire protocol involves handshaking commands, data transmission as well as an error correction based on post-processing Hamming 7,4 code. The module is developed using \lv , a proprietary software development IDE from National Instruments Inc. USA}

{\bf keywords: Polarization Shift Keying, \lv, Handshaking, Hamming code.}

{\bf PACS : 7.05. -t, 07.05.Hd, 07.05.Kf, 42.79.Sz }

\section{Introduction}
PolSK involves encoding the message bits with polarization state of a light pulse during transmission. The bits 0 and 1 are respectively mapped to two orthogonal states of polarisation. PolSK has several advantages since light can be decomposed into several sets of mutually orthogonal polarisations, such as Vertical/Horizontal,  45$^\circ$/135$^\circ$ or RCP/LCP combinations. In each of this case, the two encodings are mutually orthogonal, in the sense that a light polarised in one state will show zero for measurement for other polarization. This results in an unambiguous measurement, except in presence of noise. This also becomes particularly useful in multibit-per-symbol transmission \cite{benedetto}, although the different basis are not mutually exclusive and hence can provide ambiguity. 

We have developed an integrated software module to control and automate the above protocol using \lv{}. The program involves all relevant modules necessary to be used in PolSK communication. Although it is developed with our particular laboratory setup in mind, it is independent of the hardware involved and can be adopted with any other similar or compatible hardware. 

\section{The LabVIEW Program}
Labview is a graphical programming interface developed and distributed by National Instruments Inc. USA \cite{labview, elliot}. This has two distinct advantages over other programing environments - (i) the graphical method of programming makes it easier by removing the need to remember the code words, (ii) it has built-in modules to interface many different types of hardware units and (iii) can create an executable binary version, which can run on a different computer without installing Labview, although this facility is  available only on the professional versions. The ease of use and availability of extensive built-in modules has made \lv{} very popular in case of laboratory automation as well as controlling communication protocol \cite{labview,elliot, chun-yan}. While many of the earlier work uses \lv{} options to control TCP/IP and built-in communication protocols, we use a control program through  DAQ card, which makes the system independent of hardware, i.e., the hardware consisting of diode lasers or APD's can easily be replaced without any modification of the software. 

We have two independent parts of the code - the transmitter and the receiver part, each running on two independent computers. The synchronization between these two codes is explicitly obtained by the Labview code and hence does not demand identical clock speeds or memory for these two computers. In other words, the labview code is independent of the hardware parameters of either of the computers. 

The code described here is particularly designed for use in our experimental setup, although it can easily be used for any other schemes as it is,  or at best with  very little  modification. Our setup is described in detail in an earlier communication \cite{soorat} and will be very briefly recounted here for sake of completeness. 

\subsection*{The Transmitter}
\begin{figure}[!h]
\centerline{\includegraphics[scale=0.3]{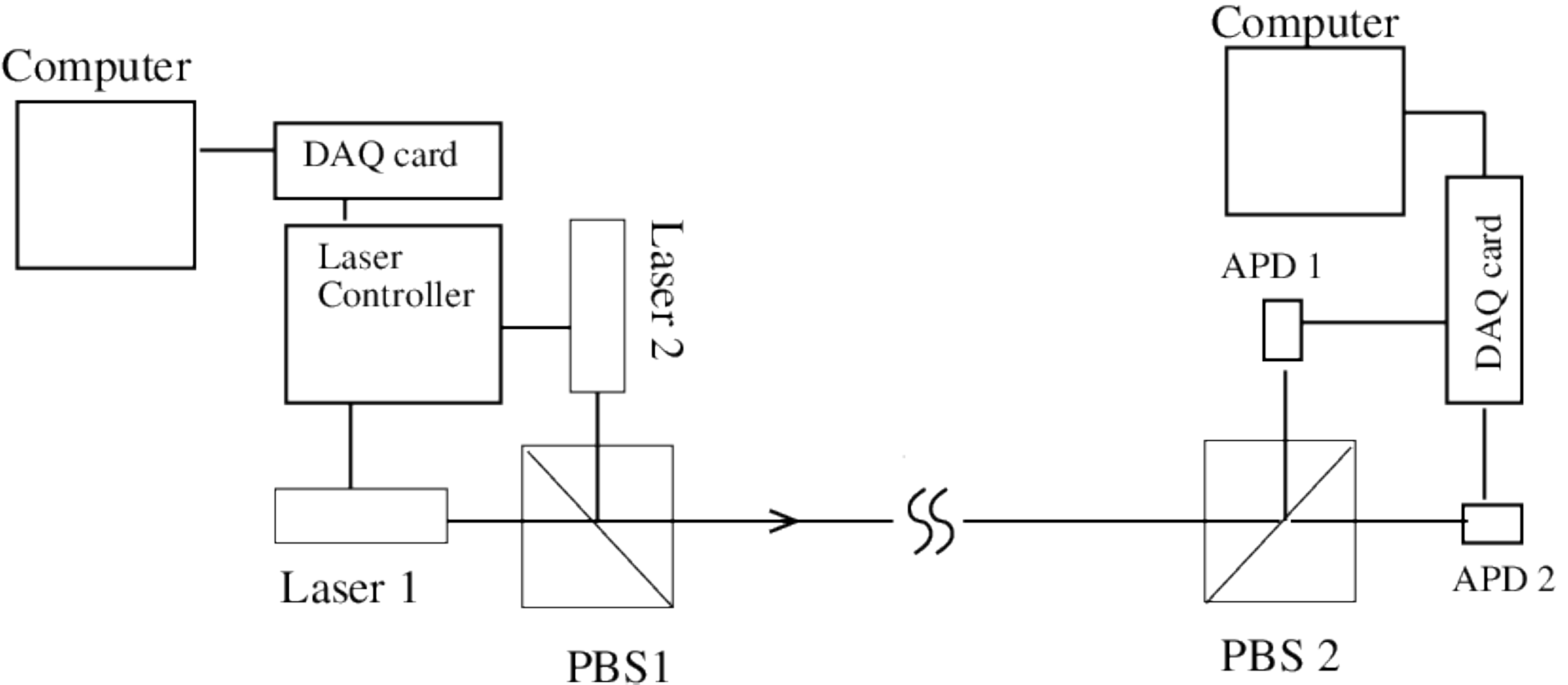}}
\caption{Schematic of the communication setup. Laser 1 and 2 are VCSEL lasers. PBS are polarizing beamsplitters. APD are Avalanche Photo Diode Modules}
\label{setup}
\end{figure}

The transmitter  consists of two VCSEL lasers, operating at 780 nm placed about a Polarizing beam splitter (PBS) as shown in figure \ref{setup}

The lasers and the PBS are arranged in such a way that vertically polarized part of  light from laser 1 and horizontally polarised part of light from laser 2 are coupled into the communication channel. The lasers are controlled by the computer through a DAQ card, which in turn fires the laser controller. Laser 1 would be switched on if the message bit is 0 and Laser 2 would be switched on if the message bit is 1. This would require the Labview code to encode as

\begin{center}
\begin{tabular}{c|c}
bit & encoding \cr \hline
0   & 01 \cr 
1   & 10 
\end{tabular}
\end{center}

The Labview code achieves this by operating a transformation 
$$k \rightarrow  2^k \xrightarrow{Binary} {\rm encoded~number }$$
where $k$ is the message bit and the encoded number be as per table above. In addition, a clock pulse, either another VCSEL in case of an all-optical networking, or a pulse transmitted over a wire in case of a hybrid connection could be used. We have tried both and the same software accounts for either of the method. With the clock pulse, the encoding scheme becomes 

\begin{center}
\begin{tabular}{c|cc}
bit & encoding \& clock \cr \hline
0   & 011 \cr 
1   & 101 \cr
no bit & 000
\end{tabular}
\end{center}

\begin{figure}[!h]
\centerline{\includegraphics[scale=0.3]{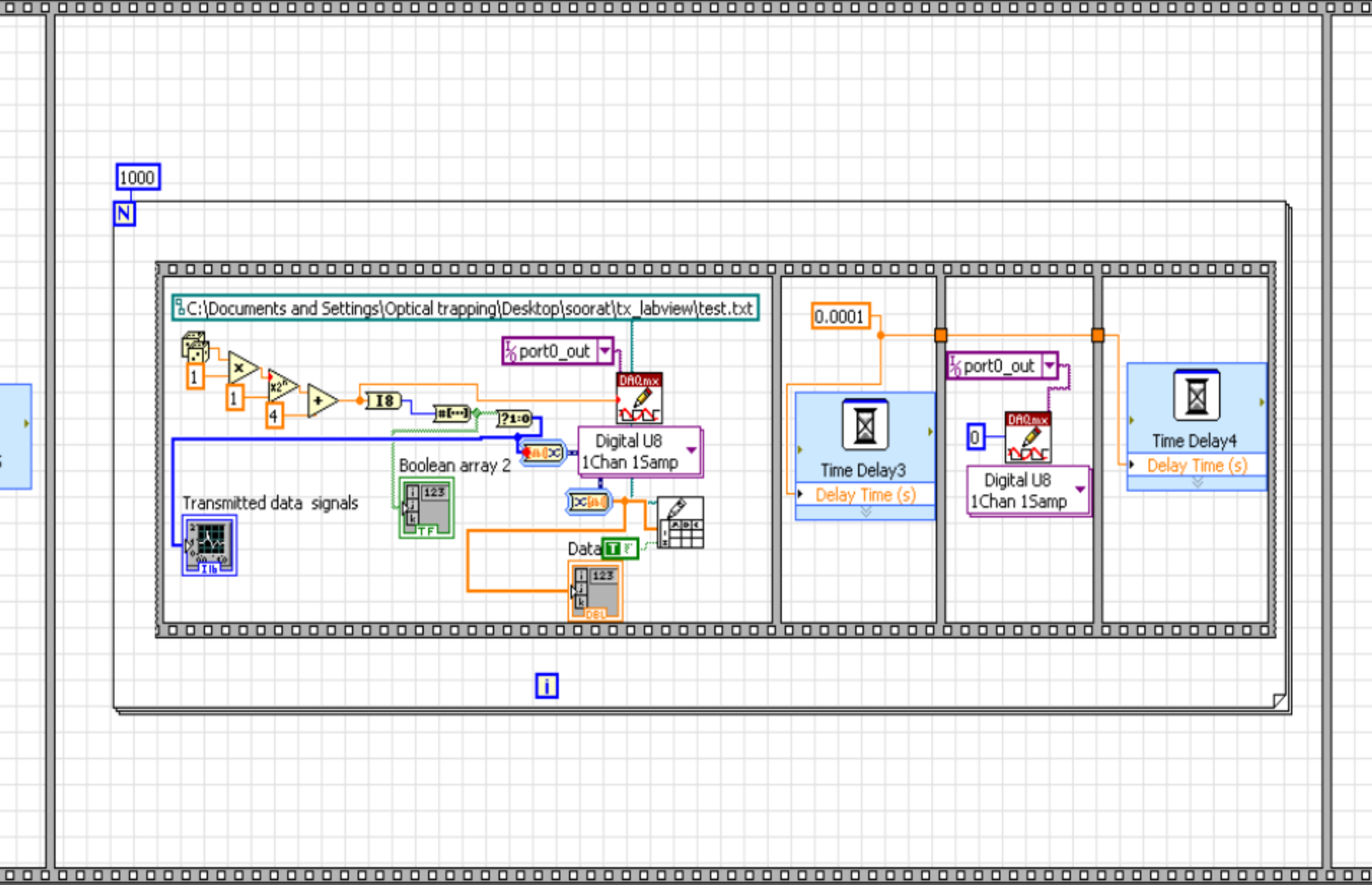}}
\caption{initial module of the transmitter. Generates a random sequence of 0's and 1's, converts it into a relevant format and writes  it onto the digital port of the DAQ card using the DAQmx module of LabVIEW.} 
\label{random_number}
\end{figure}

The need for this clock pulse is explained the receiver section. Figure (\ref{random_number}) shows the initial module of the transmitter which generates a random sequence of 0's and 1's, converts it into a relevant format and writes  it onto the digital port of the DAQ card using the DAQmx module of LabVIEW. This module was later replaced by the one which reads actual data from a saved file on the disk and similarly sends it to the DAQ card's output, which is shown in section \ref{transmitter} 

\subsection*{Receiver}

The receiver consists of another PBS whose output is incident on two Avalanche Photo Diodes APD1 and APD2. The APD's used in our setup was PCD-mini 200 from SenSL. Any other SPC module which operates in Gieger mode and provides TTL pulses for each incident photon would work equally good. The TTL pulses from the APD module are connected to counter pints of the DAQ card NI - PCI - 6320. This card has 4 counter inputs, each with 32 bit resolution and a rate of 25 MHz with external clock.     The receiver part of the Labview code is designed to obtain these TTL pulses during the on-time of the clock pulse and total them. The code resets the total number each time the clock pulse is off. This aspect required a certain round about method within the code since the original counter-handling aspect built into the Labview does not contain the reset feature, which is otherwise very important for our protocol. 

\subsection*{Operation of the protocol}
\begin{figure}[!h]
\begin{center}
\subfigure[]{\includegraphics[scale=0.3]{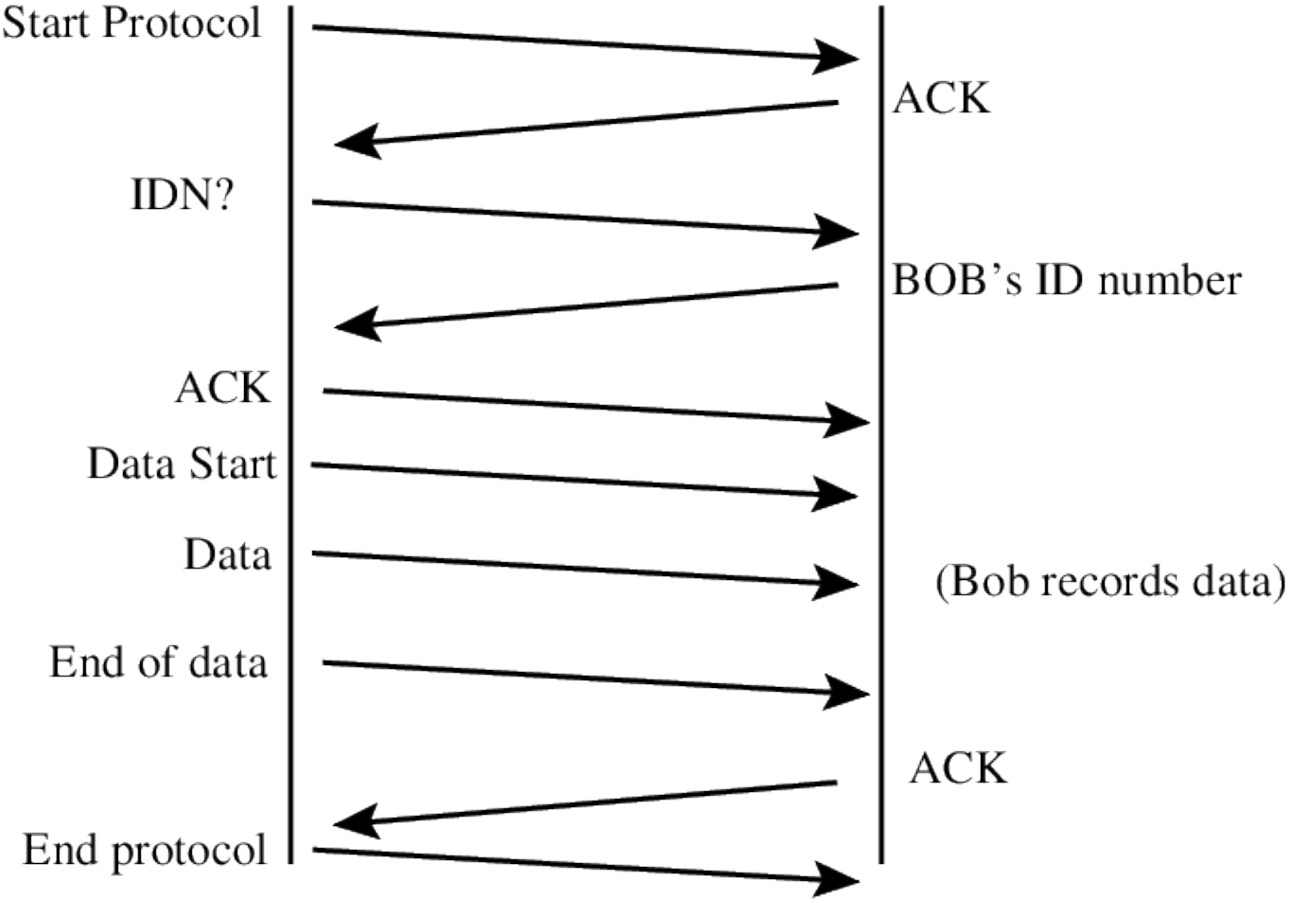}\label{protocol}}\subfigure[]{\includegraphics[scale=0.2]{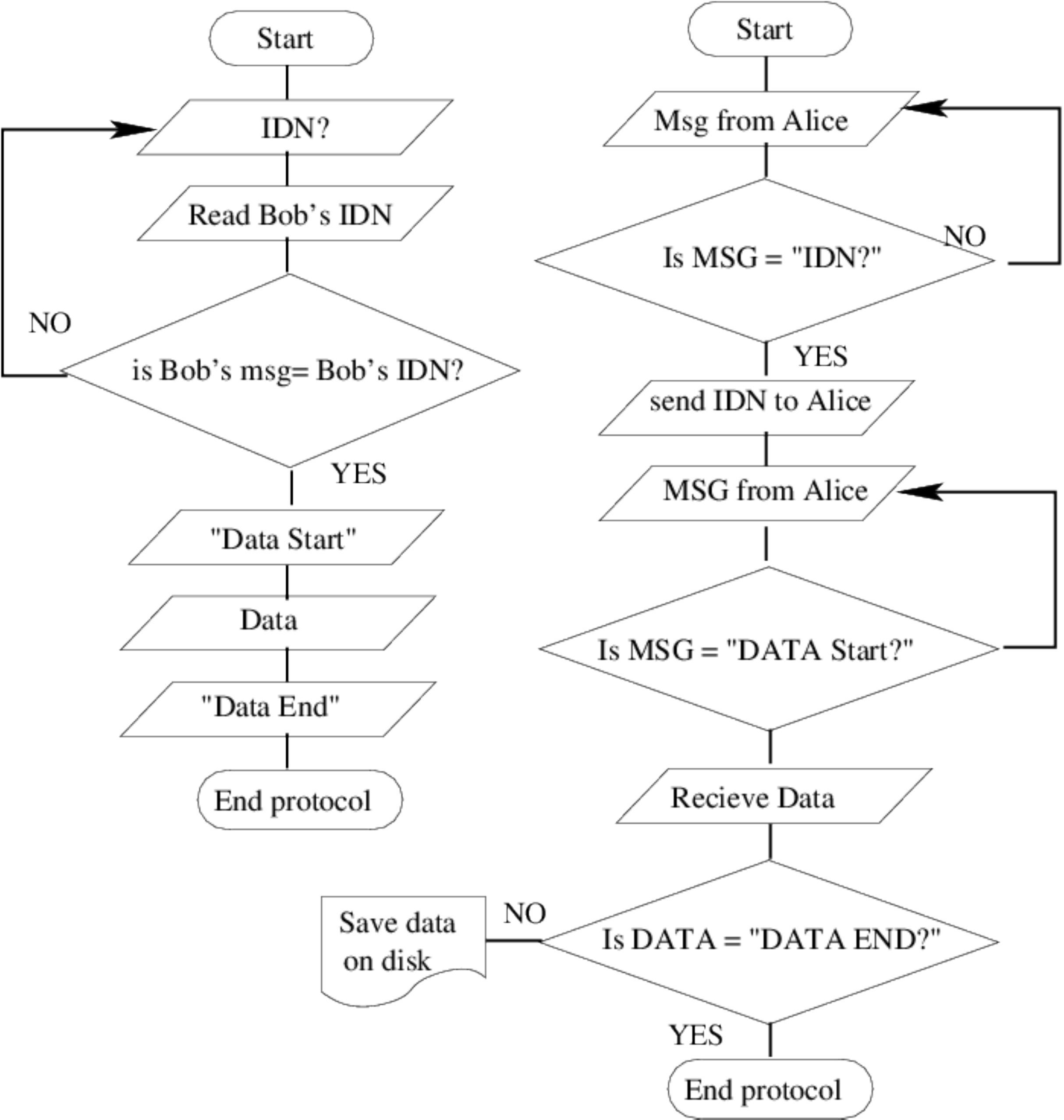}\label{flowcharts}}
\caption{(a) Elements of the protocol and (b) Corresponding flowchart of decision making}
\end{center}
\end{figure}

The generic aspects of the protocol is graphically represented in figure (\ref{protocol}). As per this, the steps involved are (i) Alice opens up the protocol with a wake up call (ii) Bob acknowledges (iii) Alice asks Bob to identify himself - so as to ensure that the data is given only to authorized receiver. (iii) Bob answers with a pre-agreed ID number. (iv) Alice compares this with the one stored in her computer and if it matches the protocol proceeds. She acknowledges it (v) Then Alice starts a Data Start code and proceeds to transmit all the data. (vi) Bob stores the data on his computer (vii) Alice concludes transmission with an EoD code. (viii) Bob acknowledges receipt of all data (vi) Alice concludes the protocol with End-of-Protocol code. 

The corresponding flow chart for the transmitter and receiver are given in figure \ref{flowcharts}. They represent the steps (i) through (viii) described above. Two individual LabView codes are 

\begin{figure}[!h]
\begin{center}
\includegraphics[scale=0.3]{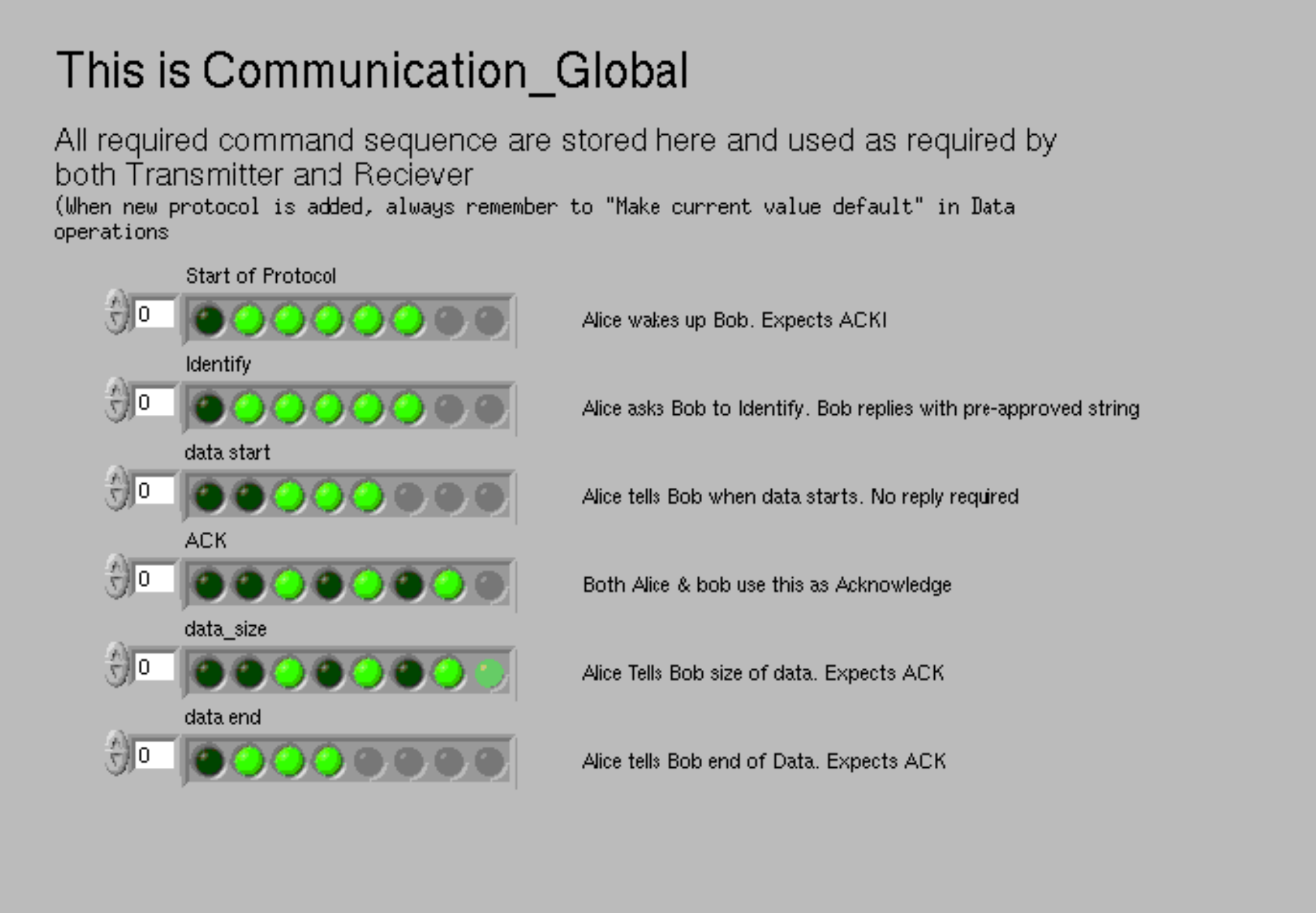}
\caption{Front panel of the Global VI, showing all the codes for handshaking}
\label{global}
\end{center}
\end{figure}

The protocol makes use of a set of standard commands which are used by both Alice and Bob, such as codes for Acknowledgment (ACK), Start of Data etc. Since these codes are a one-time standard, shared both by Alice and Bob, and any other parties involved, a we create a set of  global variables for this purpose.  The entire set can be shared as it is by all parties. The set of codes are shown in figure (\ref{global})

These codes are same for both Alice and Bob, and they pick appropriate global variable from the Global VI. A typical set of global codes would look like

\begin{table}[!h]
\begin{tabular}{lcl}
{\bf Mnemonic} & {\bf Binary code} & {\bf Description}  \cr \hline
Protocol Start & 0011 1110 & Alice wakes up Bob \cr
Identify & 0010 1110 & Alice asks Bob to identiy -  \cr
  &      & Bob answers a  \cr
  &      & pre-approved Identity code  \cr
Data start & 0001 1100 & Alice tells Bob when \cr
  &      &  data starts.  \cr
  &      & Bob starts recording data \cr
ACK & 0101 0100 & Acknowledge  \cr
Data size {\it n} & 1101 0100 & {\it n} is size of data in kbits \cr
Data end & 0000 1110 & End of Data (by Alice) \cr
Protocol End & 1111 1111 & Final closing of protocol
\end{tabular}
\caption{Typical set of codes for the global handshaking}
\end{table}

\begin{figure}[!h]
\begin{center}
\subfigure[LabView module to transmit the global VI codes]{\includegraphics[scale=0.2]{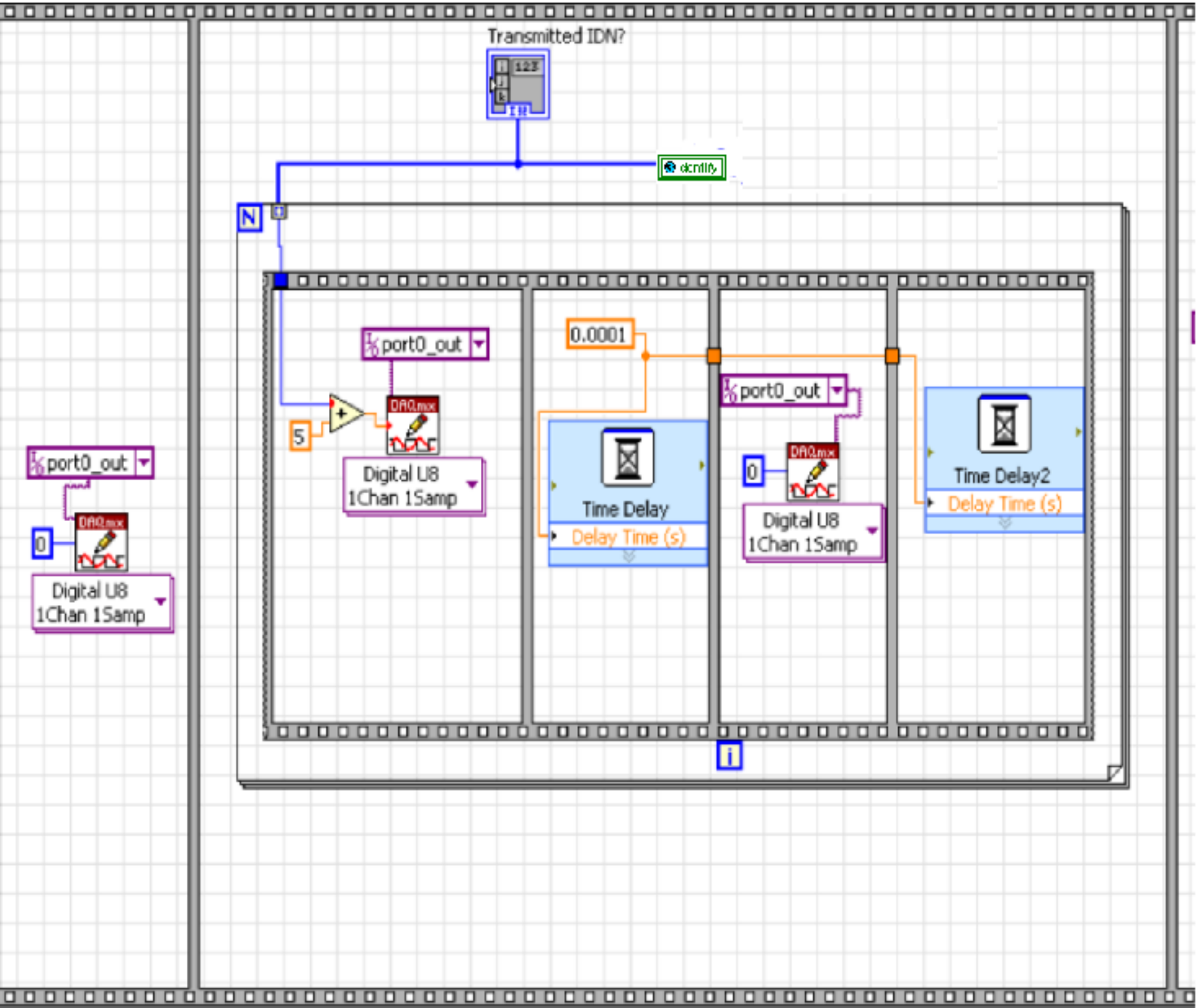}\label{transmit_global}} \hspace{1cm}\subfigure[LabView module to receive the global VI codes from Bob and compare with the stored bob's IDN ]{\includegraphics[scale=0.2]{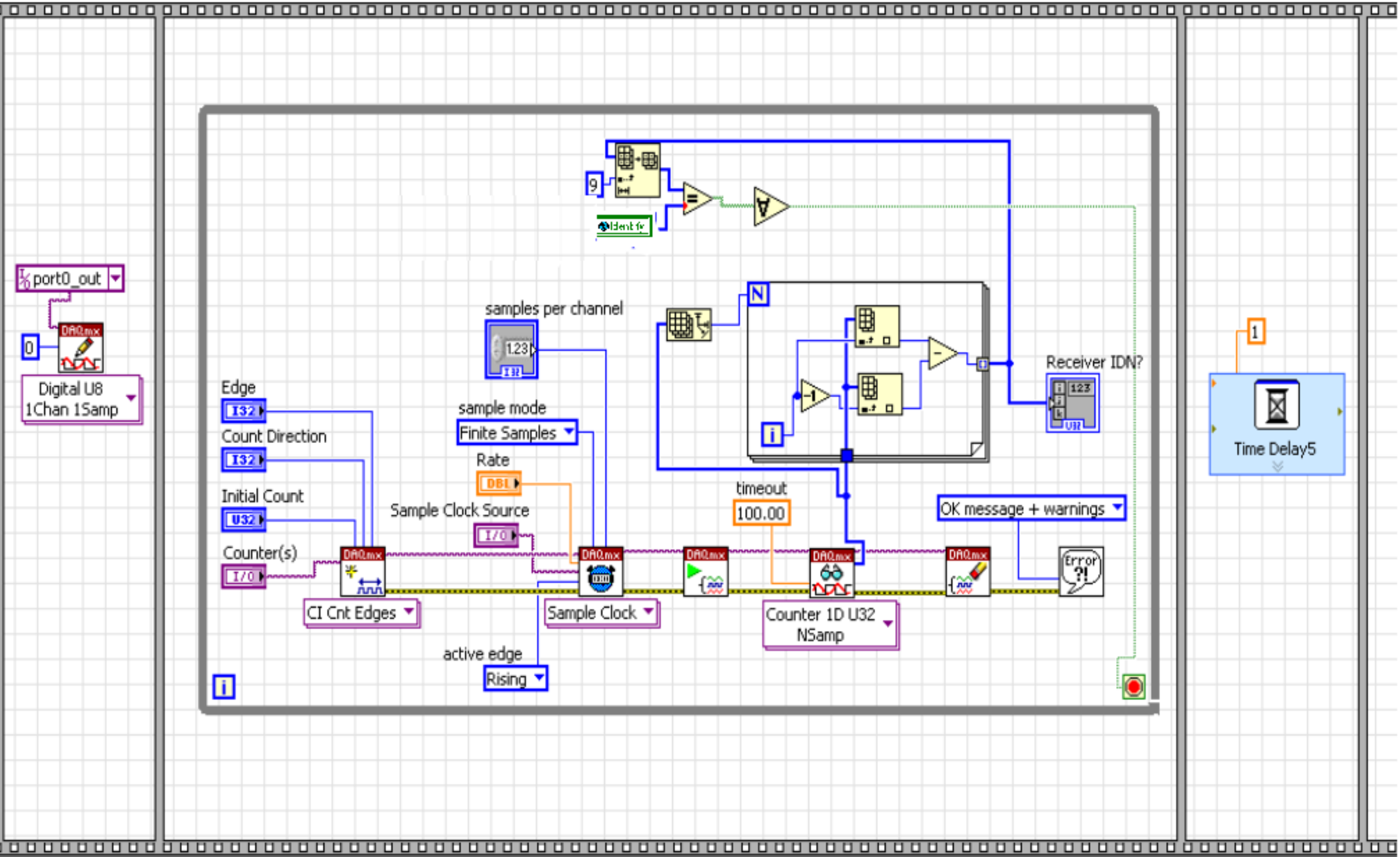}\label{receive_global}}
\end{center}
\end{figure}

Handshake protocols are shown in figures fig \ref{transmit_global} and \ref{receive_global}. These two modules are for transmission and receiving 
Figure \ref{transmit_global} shows the LabVIEW module for the handshake protocol. These modules exchange  appropriate words from the Global VI between transmitter and receiver. 

\section{Transmitter module\label{transmitter}}
\begin{figure}[!h]
\begin{center}
\includegraphics[scale=0.4]{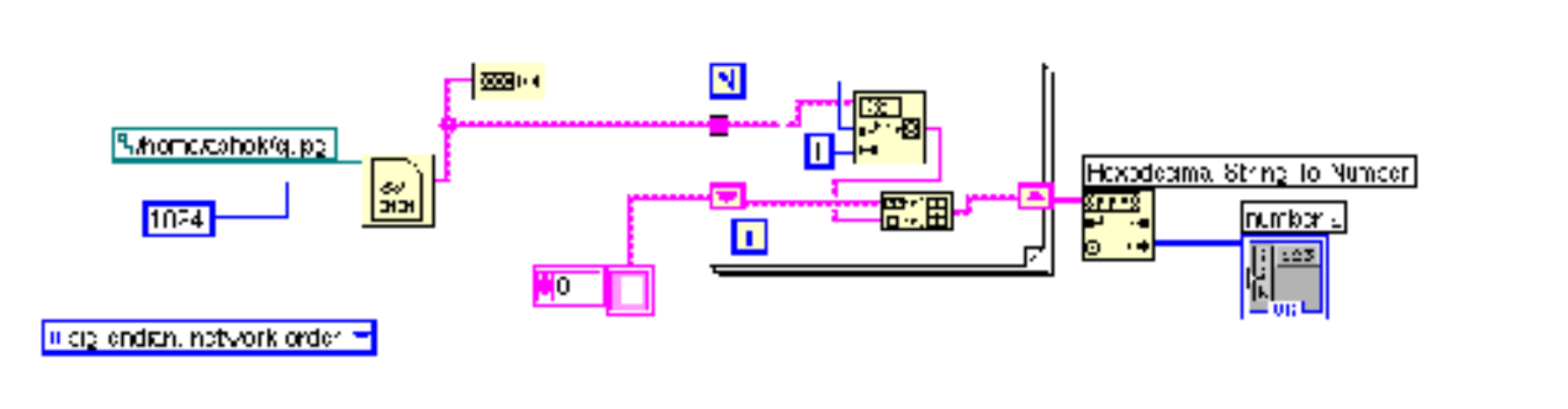}
\caption{Labview Code to read a binary file, in this case an image and extract individual bits for serial transmission}
\label{read_binary}
\end{center}
\end{figure}

The \lv{} module for transmitter is shown in \ref{read_binary}. This program reads a binary file from the hard disk and convert it into individual bits for transmission through serial transmission. These bits are then converted into 8 bit word and then sent to write-to-port module of  DAQ card. 

\subsection{Receiver}
The LabVIEW component for the receiver runs on an independent computer, recording data from the two APD modules as shown in schematic \ref{setup}. The program is started independently of the transmitter program and therefore at the beginning has a waiting loop with continuously analysing signals received at the input. If the signal corresponds to previously agreed handshaking commands from Alice, only then the program goes to the next step. The all optical version of the module receives and analyses signal. The complete module is shown in figure \ref{full_receiver}. It contains four subunits, which are described below and shown separately in following figures for sake of clarity.

\begin{figure*}
\includegraphics[width=\textwidth,height=4cm]{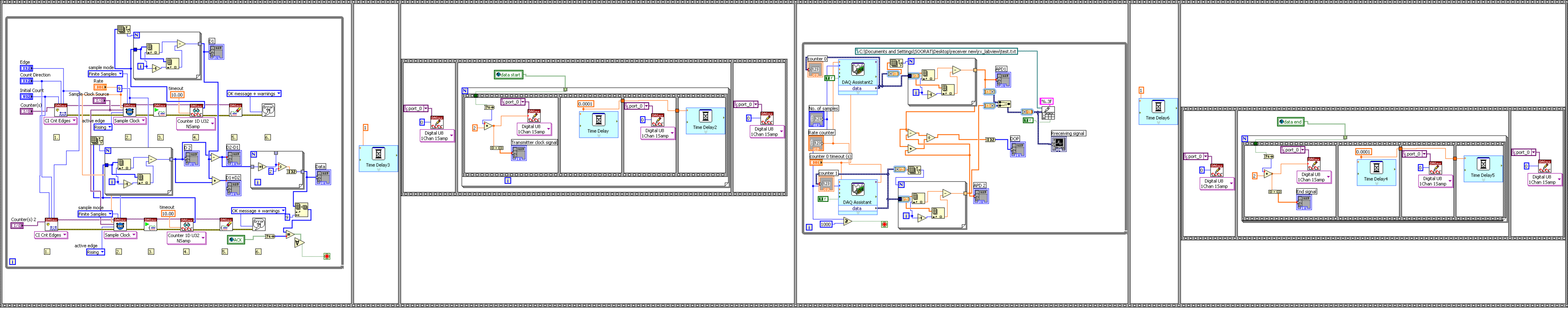}
  \caption{The complete receiver module}
  \label{full_receiver}
\end{figure*}

The first module is shown in figure \ref{wait_loop}. The two counters are initialized (top and bottom of the figure) as well as the received signal is compared for 'Identify' code. The identification of Bob is the merely ensure that Alice is transmitting the message to only authorized receiver and not to any one else.

\begin{figure}[!h]
\centerline{\includegraphics[scale=0.3]{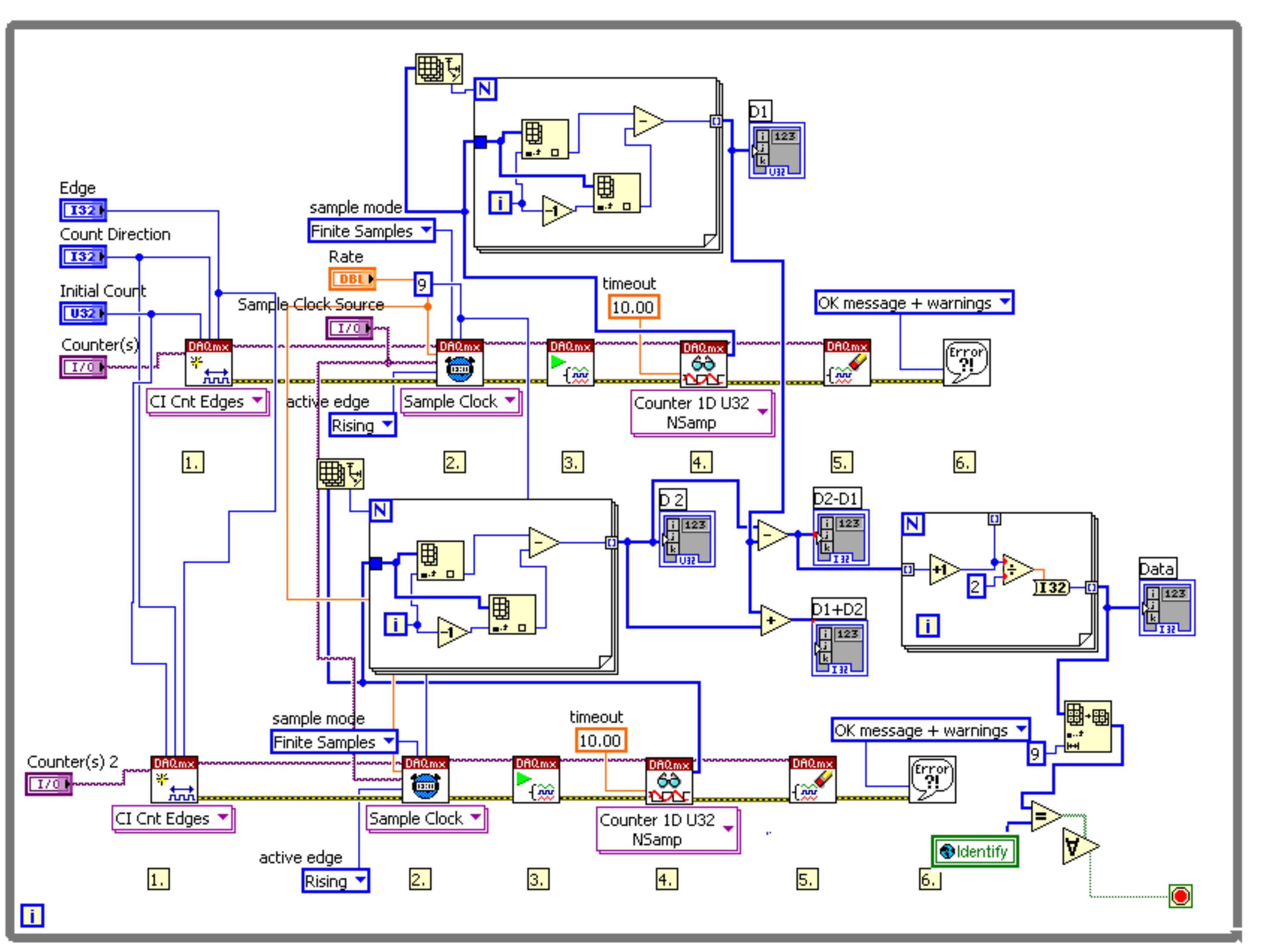}}

\caption{First module of the Receiver. It initialises relevant parameters of DAQ and also waits in a `while' loop until it receives an 'Identify' command from receiver.}
\label{wait_loop}
\end{figure}

As mentioned in previous section, the module is built around NI DAQ card PCI 6320 but in reality uses LabVIEW modules which are hardware independent and hence can be used with any other card with similar features. The APD modules provides TTL signals for every photon that is incident on them, which are fed to the counter inputs of the DAQ card. The LabVIEW module follows the steps (i) Create a Counter Input channel to Count Events. The Edge parameter is used to determine if the counter will increment on rising or falling edges. (ii)Call the DAQmx Timing VI (Sample Clock) to configure the external
sample clock timing parameters such as Sample Mode, Samples per Channel, and Sample Clock Source. The Edge parameter can be used to determine when a sample is taken. (iii) Call the Start VI to arm the counter and begin counting. The counter
will be preloaded with the Initial Count. (iv)  For finite measurements, the counter will stop reading data when the
Samples per Channel have been received. (v) Call the Clear Task VI to clear the Task. (vi) Use the pop-up dialog box to display an error if any.

The program takes input from two APD's on two counter channels, totals the TTL signals obtained within till the edge detector detects fall of the clock pulses. Since the modules are not equipped with intermittent resetting of the counter, the counts within a clock pulse duration is obtained by subtracting old total from new total. The software also computes 'State of Polarization' as given by  \cite{soorat}
\begin{equation}
S=(APD_1 - APD_2)/(APD_1 + APD_2)
\label{SOP}
\end{equation}
APD$_{1,2}$ indicates counts on respective APD's within the clock pulse duration. If the value of $S$ is positive, the program assigns data to 1 and if $S$ is negative, the data is assigned to 0. As explained in reference \cite{soorat}, this differential method of measurement provides a higher threshold against depolarization noise due to atmospheric effects.

\begin{figure}[!h]
\centerline{\includegraphics[scale=0.25]{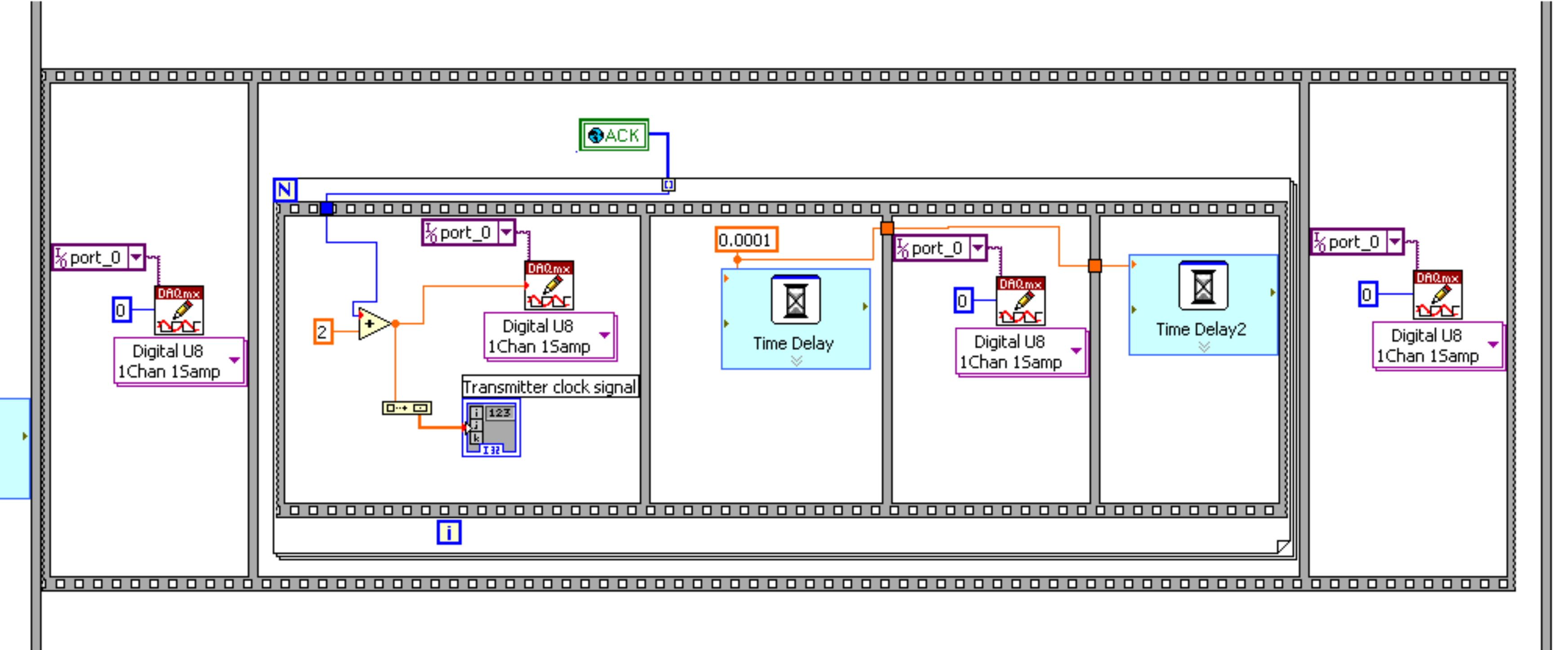}}
\caption{Second module of the Receiver. The module sends an acknowledge signal to Alice and proceeds further.}
\label{second_module}
\end{figure}

The second module takes care of handshaking, in particular that of  sending ACK signals to Alice. The third module is the main subunit which computes SoP as per equation \ref{SOP} and also saves the data onto a file for further processing. 

\begin{figure}[!h]
\centerline{\includegraphics[scale=0.3]{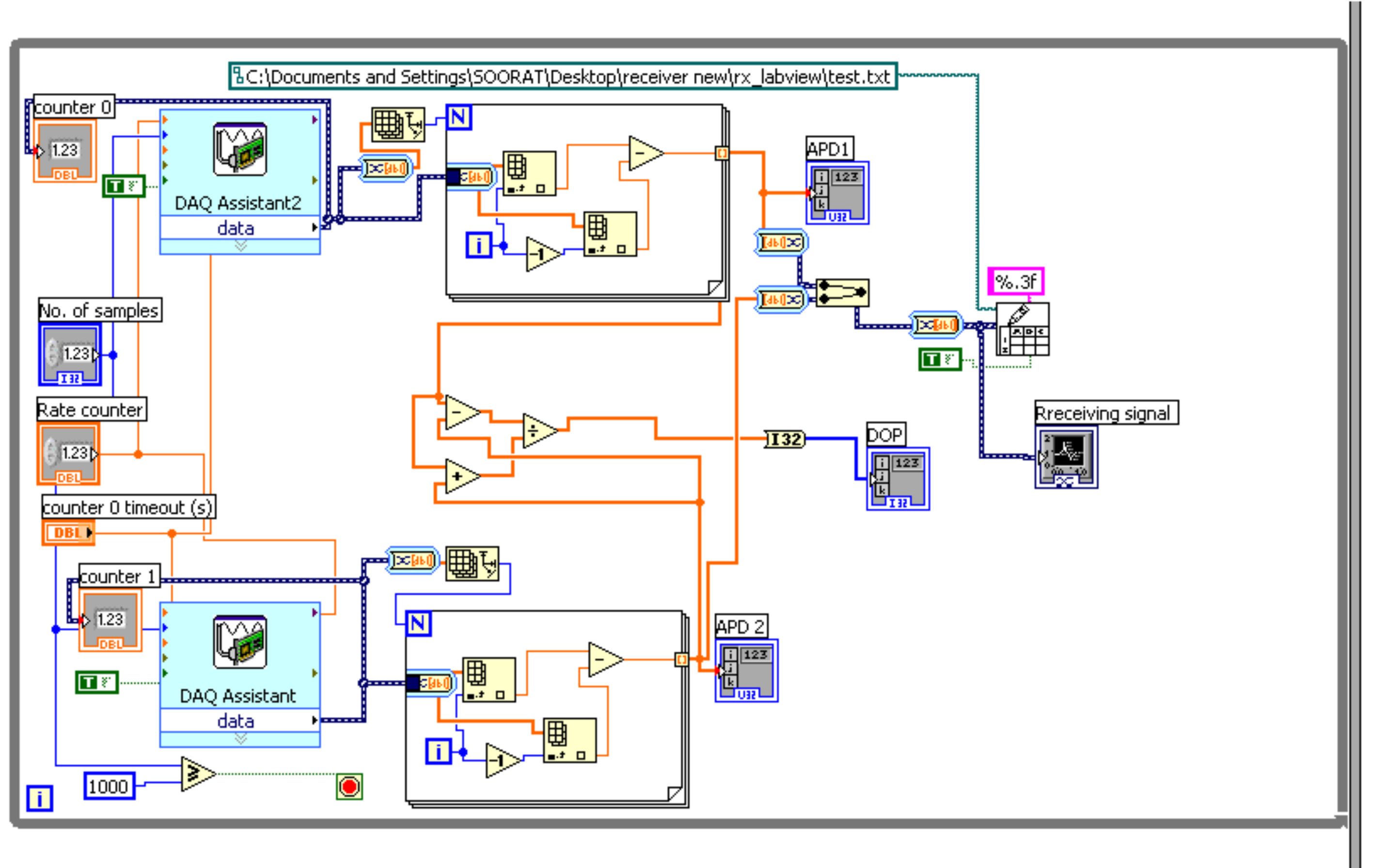}}
\caption{This is the main unit of receiver, which takes the counts from two counters and computes State of Polarization (labled as Degree of Polarization  - DoP on module) and also saves this data on a file on disk. }
\label{compute_sop}
\end{figure}

\begin{figure}[!h]
\centerline{\includegraphics[scale=0.25]{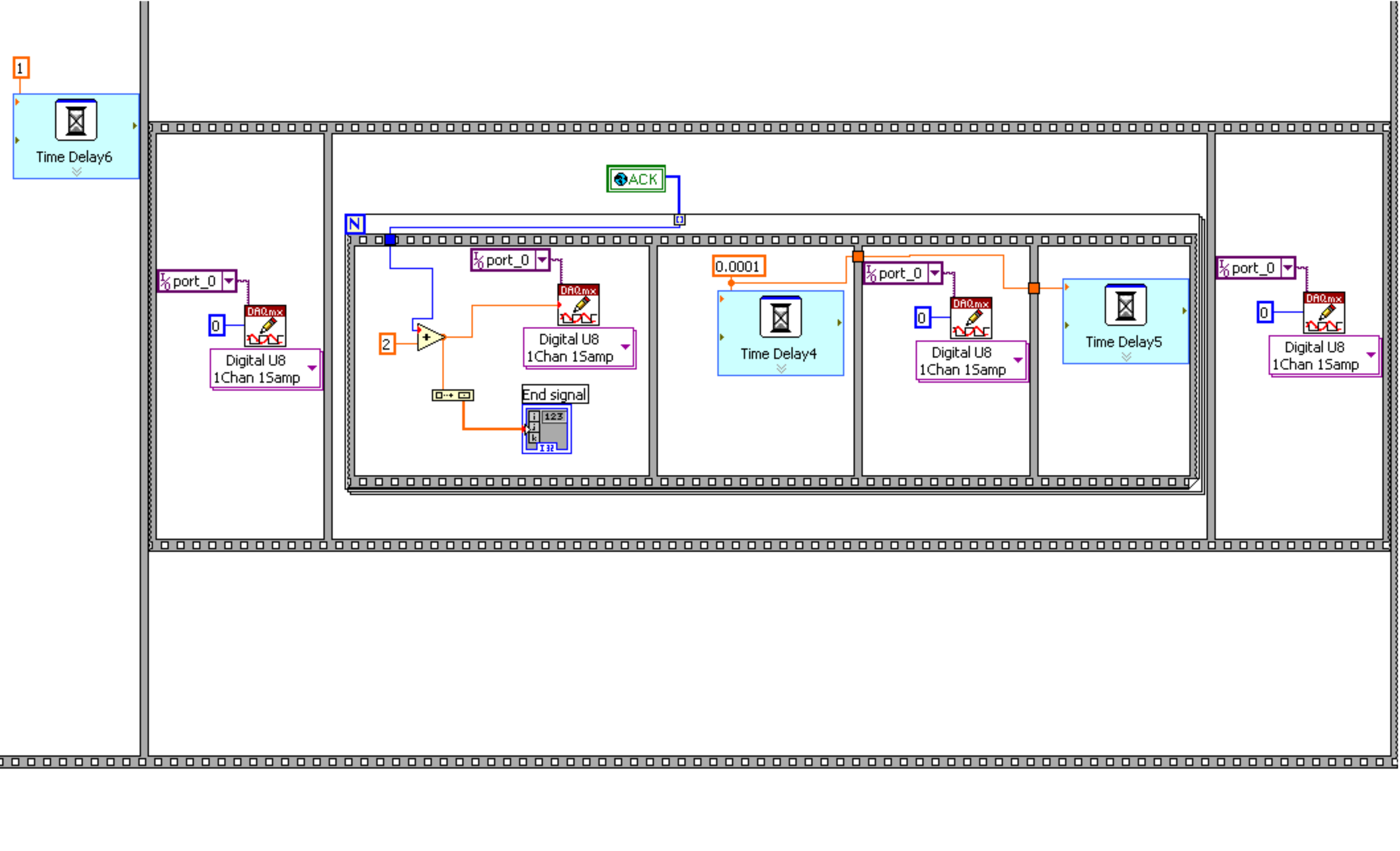}}
\caption{Final module of the Receiver. Saves all the data and sends an ACK to Alice}
\label{wait_loop}
\end{figure}

\section{Error Correction using Hamming codes}
Hamming code \cite{hamming} is one of the industry standard algorithms to detect and correct bit flip errors during transmission. The most practical configuration is the (7,4) mode, wherein three parity bits ($p_1, p_2$ and $p_3$) are added to four data bits ($d_i,~i=1,2,3,4$), adding upto 7 bits. 

\begin{eqnarray}
p_1 &=& d_1 \oplus d_2 \oplus d_4 \cr
p_2 &=& d_1 \oplus d_3 \oplus d_4 \cr
p_3 &=& d_2 \oplus d_3 \oplus d_4 
\end{eqnarray}

While in normal circumstances the parity bits are computed and interspersed with the data so as to make a 7 bit word as 
$p_1, p_2, d_1, p_3,$ $d_2, d_3, d_4$. However, since we started working with random numbers initially, we adopted a method wherein the data are sent at first and the parity bits are computed and transmitted later. This method was particularly adopted for use in a future use of  Quantum Key Distribution protocol \cite{gisin} wherein the data bits would be sent through quantum channel while the parity through the classical channel. However, the mathematics related to the computation and use of the syndrome set was identical whether the parity bits were transmitted interspersed with data bits or otherwise. 

The \lv{} code for this part consists of the transmitter part computing the relevant parity bits and create the Generator matrix G. The receiver code has modules to use this matrix G, identify the error as per standard methods and then correct them. 

\begin{figure}[!h]
\centerline{\includegraphics[scale=0.3]{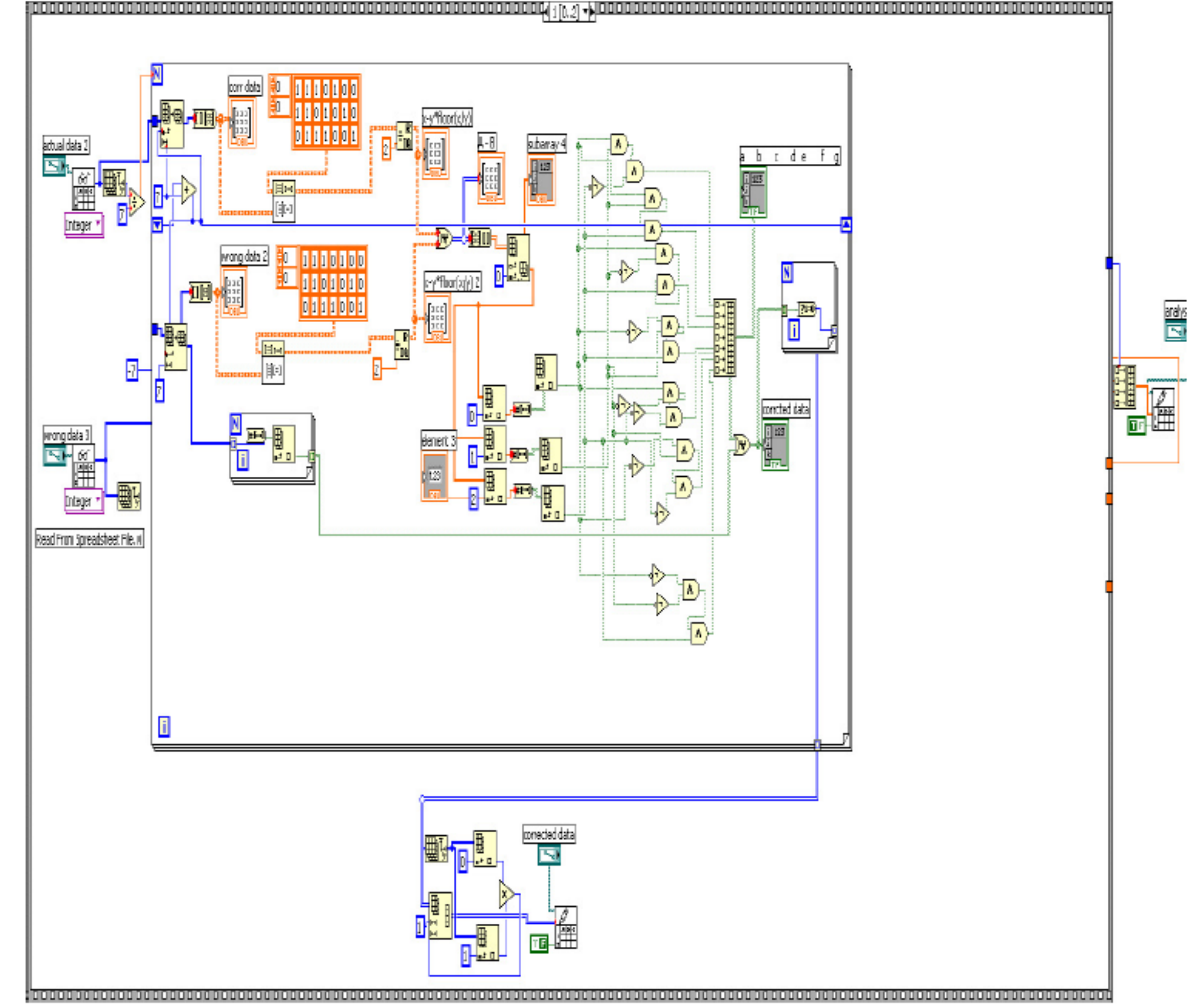}}
\caption{Receiver module of the Hamming correction code to read the generator matrix G and operate it on the data}
\end{figure}

This protocol is valid to correct any single bit flips within each set of four data bits. 

\section{conclusion}
We have prepared an integrated LabVIEW program for use of free space optical communication system using Polarization Shift Keying with Binary coding. The program is designed to control two diode lasers, each providing light polarized in orthogonal directions, initially mapped to bits 0 and 1. Corresponding light polarization is measured at the receiver, with the help of a Polarizing Beam Splitter (PBS) and a pair of Avalanche Photo diodes (APD). This measurement is in form of a `State of Polarization', which takes into account any polarization scrambling during traverse through atmosphere. The differential method adopted provides a higher  threshold against noise. The \lv{} program presented here integrates all these aspects as well as all the relevant handshaking commands. It also includes a (7,4) Hamming code error correction protocol, but with a post process option. 
 
Two  main advantages of \lv{} is that of having a graphical programming interface as well as several required modules already built-in. In this program, we exploit the digital output of a DAQ card to pulse the lasers appropriately for transmitter side and  time synchronised counter acquisition to count the pulses from an SPCD module on the receiver side. The protocol is completely integrated and contains handshaking protocols as well as the Hamming code for error correction. At the same time, it is also modular so that individual components can be corrected or replaced as required. Full professional versions of \lv{} also allow creating a stand-alone executable module, which can be run on independent computers. Future goal is to convert the present protocol for embedded versions such as FPGA or Raspberry Pi. 

\section{Acknowledgment}
This work was partially supported by Department of Information Technology, Govt. of India. RS thanks UGC-RGNF grant for fellowship.

\end{document}